\documentclass[12pt]{article}

\usepackage[totalwidth=460truept,totalheight=622truept]{geometry}
\usepackage{latexsym,graphicx,amsfonts,amssymb,amsmath,xcolor}
\usepackage[hypertexnames=false,hidelinks]{hyperref}

\def\theequation{\arabic{section}.\arabic{equation}}
\newcommand{\sect}[1]{\setcounter{equation}{0}\section{#1}}
\renewcommand{\theequation}{\thesection.\arabic{equation}}
\global\arraycolsep=1truept
\renewcommand{\theequation}{\arabic{section}.\arabic{equation}}

\usepackage[format=hang,font=small,textfont=it]{caption}

\begin{document}

\null

\vskip1truecm

\begin{center}
{\Large \textbf{High-Energy Behavior of Scattering Amplitudes}}

\vskip.6truecm

{\Large \textbf{in Theories with Purely Virtual Particles}}

\vskip1truecm

\textsl{{\large Marco Piva}}

\vskip .1truecm

{\textit{Faculty of Physics, University of Warsaw,\\ Pasteura 5, 02-093 Warsaw, Poland}}

\vspace{0.2cm}

mpiva@fuw.edu.pl
\vskip1truecm

\vskip .5truecm

\textbf{Abstract}
\end{center}

We study a class of renormalizable quantum field theories with purely virtual particles that exhibits nonrenormalizable behavior in the high-energy limit of scattering cross sections, which grow as powers of the center-of-mass energy squared and seems to violate unitarity bounds. We point out that the problem should be viewed as a violation of perturbativity, instead of unitarity, and show that the resummation of self energies fixes the issue. As an explicit example, we consider a class of $O(N)$ theories at the leading order in the large-$N$ expansion and show that the different quantization prescription of purely virtual particles takes care of the nonrenormalizable behavior, making the resummed cross sections to decrease at high energies and the amplitudes to satisfy the unitarity bounds. We compare the results to the case of theories with ghosts, where the resummation cannot change the behavior of cross sections due to certain cancellations in the high-energy expansion of the self energies. These results are particularly relevant for quantum gravity.

\vfill\eject

\section{Introduction}
\label{intro}\setcounter{equation}{0}
Nonrenormalizable quantum field theories are considered, for any practical purpose, as effective theories, namely they can be predictive only up to some energy scale. The reason for this is twofold. First, nonrenormalizability implies that infinitely many operators are generated by renormalization. Those operators have dimensions that increase with the number of loops and are multiplied by independent new couplings. Therefore, the higher the energy of the physics at play, the more the number of parameters necessary to describe it. Second, the presence of couplings with negative mass dimension leads to cross sections that can grow as powers of the center-of-mass energy squared, which makes the perturbative expansion not reliable above a certain energy scale.

However, the last property can appear also in renormalizable theories with higher derivatives. In fact, certain cancellations between tree-level diagrams lead to the same cross sections that would be obtained from the theory without the higher derivatives. If such a theory is nonrenormalizable, then the cross sections will grow as powers of the center-of-mass energy squared. A well-known example is Stelle gravity~\cite{Stelle:1976gc}, a renormalizable extension of Einstein theory. Besides the massless graviton and a massive scalar field, the theory propagates a massive spin-2 ghost responsible for the violation of unitarity. In this theory, all the tree-level scattering cross sections that involve only gravitons coincide with those of general relativity~\cite{Dona:2015tra}. This can be easily seen by means of a change of variables and writing an equivalent action that explicitly contains the fields of the scalar and spin-2 ghost~\cite{Anselmi:2018tmf}. Then, by deriving the interactions, it follows that only purely-graviton vertices, which are the same as in general relativity, can be used for tree-level diagrams with external gravitons\footnote{This property actually follows from a general theorem~\cite{Anselmi:2006yh}. See also~\cite{Dona:2015tra} for more details}. 

In recent years, a method for reconciling unitarity with renormalizability in quantum gravity has been developed, which involves transforming the ghost into a so-called \emph{purely virtual particle} (PVP) or \emph{fakeon}~\cite{Anselmi:2017yux, Anselmi:2017ygm}, that is to say a particle that can only circulate inside Feynman diagrams without ever appear as external, on-shell state. This can be implemented in any quantum field theory by means of a different quantization prescription (to suitably modify the diagrams that contain PVP), together with a projection at the level of the Fock space (to remove the PVP from the possible external state). This procedure is consistent with unitarity~\cite{Anselmi:2017lia,Anselmi:2018kgz} and give unambiguous results that are phenomenologically different from those of the original theory, where all the degrees of freedom are quantized in a standard way~\cite{Anselmi:2021hab, Melis:2022tqz}. We refer to the whole procedure as \emph{fakeon prescription}. 

Despite the fact that the theory of gravity with PVP is both unitary and renormalizable, the tree-level graviton cross sections still grow at high energies. In fact, the fakeon prescription leaves tree-level diagrams unmodified (with the exception of the removal of some delta functions, which are not relevant for the high-energy behavior). This might jeopardize the predictivity of the theory, since those cross sections become large at the Planck scale, where quantum gravity is most needed. 

In this paper we show that this is not the case and suggest that quantum gravity with PVP can be predictive above Planck scale, provided that certain resummations are performed.

In the literature, especially in particle phenomenology, the fact that a cross section grows at high energies is often interpreted as a symptom of the breaking of unitarity above some energy scale. The reason is that the so-called “unitarity bounds” appear to be violated. We argue that the unwanted behavior of the cross sections should be viewed as a violation of perturbativity, rather than unitarity. Indeed, we show that nonperturbative resummation of self energies, fixes the issue when PVP are present. In particular, we study a class of higher-derivative $O(N)$ theories at the leading order in the large-$N$ expansion. We show that, after the resummation, the cross sections in renormalizable theories with PVP decrease as inverse powers of the center-of-mass energy squared and that the amplitudes satisfy the unitarity bounds. These results are compared to the case of theories with ghosts, where the resummation does not help in improving the high-energy behavior. In doing so, we study also the case of nonrenormalizable theories, which can be cured by means of the partial resummation as well. 

Our claim is that if a theory satisfies the diagrammatic version of the optical theorem and violates the unitarity bounds, then nonperturbative techniques, such as partial resummations, can always make the cross sections explicitly unitary. Theories with PVP, as well as nonrenormalizable ones\footnote{However, in the case of nonrenormalizable theories the presence of infinitely many couplings generated by renormalization remains and, therefore, they still cannot be considered predictive up to arbitrary energies.}, fall in this category, while theories with ghosts do not, despite their renormalizability, and are truly not unitary. 

The paper is organized as follows. In~\autoref{sect:unit} we review the standard derivation of the unitarity bounds and point out the reason why their violation above some energy scale signals a lack of perturbativity there. Moreover, we introduce the effect of the fakeon prescription in the case of one-loop bubble diagrams, which are the only ones needed in our analysis. In~\autoref{sect:general} we introduce a class of scalar higher-derivative $O(N)$ models and show the strategy that we adopt to study the resummation in a given model, as well as some general properties. In particular, we show how the high-energy behavior of the self energies is affected by the different prescriptions. In~\autoref{sect:super} we study in detail a superrenormalizable model that belongs to the class of~\autoref{sect:general} and mimic Stelle gravity in some aspects. We show that, after the resummation, the cross sections decrease at high energies in the case of PVP and point out the qualitative difference with the ghost and nonrenormalizable cases. Finally, we show that in the case of PVP and nonrenormalizable theory the resummed amplitudes satisfies the unitarity bounds, while in the ghost case they are still violated. Section~\autoref{sect:concl} contains our conclusions. In~\autoref{app:reno}
we discuss the renormalizability of the models of~\autoref{sect:general} and~\autoref{sect:super}.
 
Our convention for the signature of Minkowski metric is $(+,-,-,-)$.

\sect{Unitarity, perturbativity and purely virtual particles}
\label{sect:unit}

In this section we review some aspects of unitarity and perturbativity. Moreover, we briefly recall the formulation of purely virtual particles in the context needed for this paper.

The unitarity condition on the scattering matrix $S$ is written in terms of its nontrivial part $T$ as
\begin{equation}\label{eq:opttheom}
    -i\left(T-T^{\dagger}\right)=\frac{1}{2}T^{\dagger}T,
\end{equation}
which is historically called \emph{optical theorem}. If each matrix element of $iT$ is expanded as a sum of Feynman diagrams, we can compare the terms of the same order. However, a more general set of equations, valid diagram by diagram, can be derived. In fact, in any quantum field theory the so-called \emph{cutting equations} hold by construction and state that for each diagram $\mathcal{D}$ we have
\begin{equation}\label{eq:cuteq}
    \text{Im}\left(-i\mathcal{D}\right)=-\sum_{\text{c}}\mathcal{D}_{\text{c}},
\end{equation}
where $\mathcal{D}_{\text{c}}$ are the \emph{cut diagrams}, which are obtained by means of additional Feynman rules. The details are irrelevant for the present discussion. The important point is that when the right-hand-side of~\eqref{eq:cuteq} can be interpreted as the right-hand-side of~\eqref{eq:opttheom} for each diagram, then the cutting equations represent the diagrammatic version of the optical theorem and the theory is unitary. This does not always hold, since it depends on the residues of propagators and a few other caveats due to gauge symmetries~\cite{tHooft:1973wag}. When it is true we call the set of equations~\eqref{eq:cuteq} the \emph{diagrammatic optical theorem}.

Another way of writing~\eqref{eq:opttheom} is by means of the \emph{partial wave expansion}, which is often used in the literature in order to impose certain bounds. The usual argument works as follows. Consider a matrix element $\mathcal{M}$ of $T$, which we call amplitude\footnote{The relation between $T$ and $\mathcal{M}$ is $\langle f|T|i\rangle=(2\pi)^4\delta^{(4)}(p_f-p_i)\mathcal{M}$, for some initial and final states $|i\rangle$,$|f\rangle$ with total momentum $p_i$ and $p_f$, respectively.}. To simplify the steps and quickly go to the point we choose an elastic scattering of two scalar particles with masses $m_1$ and $m_2$. The amplitude $\mathcal{M}$ is a function of the Mandelstam variables $s$ and $t$ or, in the center-of-mass frame, a function of $s$ and the scattering angle $\theta$. Therefore, the cross section reads
\begin{equation}
\sigma(s)=\frac{1}{32\pi s}\int_{-1}^1\mathrm{d}v|\mathcal{M}(s,v)|^2, \qquad v\equiv\cos\theta
\end{equation}
and the optical theorem~\eqref{eq:opttheom} turns into
\begin{equation}\label{eq:opttheomamp}
    \text{Im}\mathcal{M}(s,1)=\sqrt{\kappa(s,m_1^2,m_2^2)}\sum_{X}\sigma_X(s)\geq\sqrt{\kappa(s,m_1^2,m_2^2)}\sigma(s),
\end{equation}
where
\begin{equation}
    \kappa(x,y,z)=x^2+y^2+z^2-2xy-2xz-2yz
\end{equation}
is the K\"allen function, $\sigma_X$ is the cross section for the process where particles $m_1$ and $m_2$ go into a final state $X$, and the sum is over all possible final states. 

The amplitude can be expanded using the Legendre polynomials $P_j$ as basis\footnote{The general derivation for particles with spin is obtain by changing basis of the Fock space used to derive the matrix elements of $T$ and involves Wigner functions. For details see~\cite{Itzykson:1980rh}.}
\begin{equation}\label{eq:partialwave}
    \mathcal{M}(s,v)=16\pi\sum_{j=0}^{\infty}(2j+1)A_j(s)P_j(v), \qquad \int_{-1}^{1}\mathrm{d}vP_j(v)P_k(v)=\frac{2}{2j+1}\delta_{jk}
\end{equation}
and the cross section becomes
\begin{equation}
  \sigma(s)=\frac{16\pi}{s}\sum_{j=0}^{\infty}(2j+1)|A_j(s)|^2.
\end{equation}
Using~\eqref{eq:partialwave} in~\eqref{eq:opttheomamp} we get
\begin{equation}\label{eq:unitbound1}
    \sum_{j=0}^{\infty}(2j+1)\text{Im}A_j(s)\geq\frac{\sqrt{\kappa(s,m_1^2,m_2^2)}}{s}\sum_{j=0}^{\infty}(2j+1)|A_j(s)|^2.
\end{equation}
The inequality~\eqref{eq:unitbound1} is used to derived bounds for specific cases. A very common one is the situation where only the elastic channel is allowed and~\eqref{eq:unitbound1} becomes an equality leading to 
\begin{equation}
    \text{Im}A_{j}(s)=\frac{\sqrt{\kappa(s,m_1^2,m_2^2)}}{s}|A_j(s)|^2.
\end{equation}
Defining $\mathcal{A}_j(s)=\frac{\sqrt{\kappa(s,m_1^2,m_2^2)}}{s}A_j(s)$
we find
\begin{equation}\label{eq:unitbound2}
    |\mathcal{A}_j|\leq 1, \qquad 0\leq\text{Im}\mathcal{A}_j\leq 1, \qquad |\text{Re}\mathcal{A}_j|\leq\frac{1}{2}.
\end{equation}
All of the above is well known. Indeed, in the literature, the set of inequalities~\eqref{eq:unitbound2} is often used to verify whether or not a theory satisfies unitarity at arbitrary energies. The most common check is done by means of $s$-channel diagrams, since they do not depend on $\theta$. For example, suppose that a theory produces a tree-level $s$-channel amplitude $\mathcal{M}(s)=\lambda^2 s/M^2$, where $\lambda$ is a coupling constant and $M$ some mass scale. In this case, we have\footnote{For simplicity we consider massless particles.} $\mathcal{A}_0=A_0=\frac{\lambda^2 s}{16\pi M^2}$. From the first inequality in~\eqref{eq:unitbound2} it would follow that unitarity is violated for $s>16\pi M^2/\lambda^2$. We think that this conclusion is incorrect. The reason is that the above argument is based on two hypotheses: unitarity and perturbativity. In fact, the inequalities~\eqref{eq:unitbound2} are valid for the full amplitudes and no expansion in terms of Feynman diagrams is assumed there. While in the typical arguments used in the literature the perturbative expansion is performed and higher orders are considered smaller that the lower ones. On the other hand, the diagrammatic optical theorem makes use of the perturbative expansion, albeit in a formal sense. In fact, there is no need for the coefficients of higher-order terms to be smaller than the lower ones (as functions of $s$) to derive the cutting equations. It is sufficient to have an expansion in terms of Feynman diagrams (even formally) and local vertices. In quantum field theory perturbativity can be violated in many ways, even if the coupling constants are small and we can still perform expansion in Feynman diagrams, since the coefficients of the series that we obtain are functions of $s$. Therefore, we find confusing to state that a theory violates unitarity at some scale because certain tree-level cross sections grow as powers of $s$. Unitarity is equivalent to the diagrammatic optical theorem which, if satisfied, holds regardless the energy dependence of cross sections. If a cross section grows with the energy, so does the imaginary part of the associated diagram and they grow hand in hand without violating unitarity.

The correct statement is that if~\eqref{eq:unitbound2} are violated at tree level then either unitarity or perturbativity is violated. But if the diagrammatic optical theorem holds, it does independently on $s$. Therefore, perturbativity must be violated. This simply means that above some scale higher-order terms count as much as lower ones and they are necessary to explicitly show unitarity.

Another statement often seen in the literature is that the violation of~\eqref{eq:unitbound2} signals the presence of new physics that will eventually cure the violation above certain scale where such new physics is relevant.  
Adding new physics can be a solution, but there has to be a way within the same theory, since the optical theorem does not know anything about new physics. A theory, effective or not, is just our mathematical description and if satisfies the optical theorem then the cure of its apparent violation must rely on a nonperturbative mechanism within the same theory, without advocating any new physics.

In the next section we show some explicit examples where this is realized by means of partial resummations both in nonrenormalizable theories and in renormalizable theories with PVP. Finally, the resummation is performed also in theories with ghosts to show a true violation of unitarity.

Before to proceed, we recall some basic properties of PVP and how they are implemented in quantum field theory. 

The aim of introducing a particle that is purely virtual is to keep its contributions to renormalization, while being able to truly remove it from the set of degrees of freedom that can appear as external legs of Feynman diagrams. The main application of this idea is in theories plagued by the presence of ghosts, i.e. degrees of freedom with negative residue, such as renormalizable quantum gravity. In fact, in the case of Stelle gravity, the presence of the ghost ensures renormalizability thanks to its virtual contributions, while at the same time spoils unitarity in those situations where it can be on shell. The solution to this problem is to turn the ghost into a PVP. The procedure used to implement a PVP in a theory is made of two main steps: a prescription and a projection. The prescription consists in a different way of treating the amplitudes near the branch cuts associated to the physical production of PVP, while the projection restricts the Fock space onto a subspace where the PVP are not external states. The latter procedure is consistent thanks to the prescription. The generality of the fakeon prescription is addressed in several papers~\cite{Anselmi:2017lia, Anselmi:2018kgz, Anselmi:2021hab} and explicit formulas for one-loop diagrams, as well as their phenomenological implications, can be found in~\cite{Melis:2022tqz}. However, for the purposes of this paper it is enough to know how the fakeon prescription works in the case of bubble diagrams and we direct the reader to the above-mentioned references for the details beyond those presented here.

We define a bubble diagram by means of the integral
\begin{equation}\label{eq:bubint}
B_{ij}(p^2)\equiv \int\frac{\text{d}^Dq}{(2\pi)^D}\frac{1}{(p+q)^2-M_i^2+i\epsilon}\frac{1}{q^2-M_j^2+i\epsilon},
\end{equation}
where $M_i$ are masses, $\epsilon >0$ and $D$ is the extended dimension of dimensional regularization. For the moment we have used the Feynman prescription for both propagators, although in the case of PVP we need to use the fakeon prescription. There is a way to implement it in terms of propagators and Feynman integrals~\cite{Anselmi:2017yux,Anselmi:2017lia}. However, the easiest way is to compute the integrals using Feynman prescription everywhere and apply the fakeon procedure directly at the level of the amplitude. In the case of bubble diagrams the procedure simplifies to just removing the imaginary part of the amplitude (or the real part of the diagram). This means that if at least one of the particles of mass $M_i$ and $M_j$ in~\eqref{eq:bubint} is a PVP, then the fakeon prescription amounts to the substitution
\begin{equation}\label{eq:fakeonsubs}
    B_{ij}(p^2)\rightarrow \text{Im}\left[B_{ij}(p^2)\right].
\end{equation}
In the next sections we use the explicit form of the integral~\eqref{eq:bubint}, which reads~\cite{Bohm:1986rj,Anselmi:2020tqo}   
\begin{eqnarray}\label{eq:bubblediag}
&&i(4\pi )^{2}B_{ij}(p^2)=-\frac{2}{\tilde{\varepsilon}}+\ln \frac{
M_{i}M_{j}}{\mu^2}+\frac{M_{i}^{2}-M_{j}^{2}}{s}\ln \frac{M_{i}}{M_{j}} \nonumber\\
&&-\theta (u_{-})\frac{\sqrt{u_{+}u_{-}}}{s}\left( \ln \frac{\sqrt{u_{+}}+\sqrt{%
u_{-}}}{\sqrt{u_{+}}-\sqrt{u_{-}}}\right)+\theta (-u_{-})\theta (u_{+})\frac{2\sqrt{-u_{+}u_{-}}}{s}\arctan 
\sqrt{\frac{-u_{-}}{u_{+}}}\nonumber\\
&&+\theta
(-u_{+})\frac{\sqrt{u_{+}u_{-}}}{s}\left( \ln \frac{\sqrt{-u_{-}}+\sqrt{-u_{+}}}{\sqrt{-u_{-}}-\sqrt{%
-u_{+}}}-i\pi \right) ,
\end{eqnarray}%
where $2/\tilde{\varepsilon}=2/\varepsilon-\gamma_E+2+\ln 4\pi$, $\gamma_E$ is the Euler-Mascheroni constant, $\varepsilon=4-D$, $\mu$ is the renormalization scale and
\begin{equation}
u_{\pm }=(M_{i}\pm M_{j})^{2}-p^{2}.
\end{equation}%
Moreover, it is useful to introduce\footnote{No summation over repeated indices.}
\begin{equation}\label{eq:Btau}
    B_{ij}(p^2,\tau)\equiv \mathcal{T}_{ij}(\tau) \text{Re}B_{ij}(p^2)+i\text{Im}B_{ij}(p^2)
\end{equation}
where the matrix $\mathcal{T}_{ij}(\tau)$ is 1 if both particles $i$ and $j$ are standard, while it is $\tau\in\mathbb{R}$ if at least one particle among $i$ and $j$ is purely virtual. In order to truly reproduce the fakeon prescription, $\tau$ should be zero. However, in order to keep track of the effect of the different prescriptions, we keep $\tau$ nonvanishing and set it to 0 only at the end of the computations.

Removing the real part of $B_{ij}$ comes from the requirement that a PVP cannot be produced on shell. In fact, viewing the bubble diagram~\eqref{eq:bubblediag} as a function of the complex variable $p^2$, there is a branch cut on the real axis for $p^2\geq (M_i+M_j)^2$, which corresponds to an imaginary part of the amplitude. The threshold $p^2=(M_i+M_j)^2$ represents the value of $p^2$ above which a couple of particles $M_i$ and $M_j$ can be produced on shell. Therefore, if at least one of them is purely virtual, such imaginary part of the amplitude must be set to zero, hence the choice $\tau=0$. As mentioned above, setting $\tau=0$ is consistent with the fakeon projection and the degrees of freedom removed in this way cannot be generated back by quantum corrections.

The explicit formulas for the generalization of~\eqref{eq:fakeonsubs} in the case of triangle and box diagrams can be found in~\cite{Melis:2022tqz}, while we refer to~\cite{Anselmi:2021hab} for the rules at all orders.

\sect{General strategy}
\label{sect:general}
In this section we consider the lagrangian of a general scalar theory with higher derivatives, both in the kinetic and interaction terms, encoded in some polynomial functions. For the moment we do not focus on the renormalizability of the model, as some choices of the functions might lead to a nonrenormalizable theory. We choose a renormalizable example in the next section and discuss general renormalization properties in~\autoref{app:reno}. Here we show the strategy that we use to study the resummation of bubble diagrams and show how the fakeon prescription can affect their high-energy behavior. In order to justify the resummation, we consider a $O(N)$ theory and perform the large-$N$ expansion. The lagrangian reads
\begin{equation}\label{eq:generallagra}
\begin{split}
    \mathcal{L}(\varphi)=& \ \frac{1}{2}\partial_{\mu}\varphi^a F_n\left(\frac{-\square}{M^2}\right)\partial^{\mu}\varphi^a-\frac{1}{2}m^2\varphi^a F_n\left(\frac{-\square}{M^2}\right)\varphi^a-\frac{1}{8}\varphi^2 G_r\left(\frac{-\square}{M^2}\right)\varphi^2\\[2ex]
    =& \ \mathcal{L}_{\text{kin}}(\varphi)-\frac{1}{8}\varphi^2 G_r\left(\frac{-\square}{M^2}\right)\varphi^2, \qquad \varphi^2\equiv \varphi^a\varphi^a, \qquad a=1,\ldots N,
    \end{split}
\end{equation}
where $F_n(z)$ and $G_r(z)$ are polynomials of degree $n$ and $r$, respectively, and $M$ is a mass parameter. We rewrite the interaction term by introducing two auxiliary fields $\Omega$ and $\chi$ with the purpose of reducing it to a single term proportional to $\Omega\varphi^2$. Part of the old interaction is recast into the $\Omega$ propagator, which has no poles. The new lagrangian reads
\begin{equation}\label{eq:lagrachi}
    \mathcal{L}^{\prime}(\varphi,\Omega,\chi)=\mathcal{L}_{\text{kin}}(\varphi)-\frac{1}{2}\chi G_r\left(\frac{-\square}{M^2}\right)\chi+\frac{1}{2}\Omega(2\chi-\varphi^2).
\end{equation}
 It is easy to see the equivalence of the two lagrangians by solving the equation of motion for $\Omega$, which gives $\chi(\varphi)=\varphi/2$ and we have
\begin{equation}
    \mathcal{L}^{\prime}(\varphi,\Omega,\chi(\varphi))=\mathcal{L}(\varphi).
\end{equation}
However, in the rest of the paper we integrate $\chi$ out so we can explicitly write the $\Omega$ propagator. The role of the field $\chi$ is to show that the lagrangian~\eqref{eq:lagraomega} (see below) can be written as~\eqref{eq:lagrachi}, which is local. The equation of motion for $\chi$ gives
\begin{equation}
    \chi(\Omega)=G^{-1}_r\Omega,
\end{equation}
where $G^{-1}_r$ is an inverse operator of $G_r$ (which one is unimportant for our purposes, since the results do not depend on this choice). Thus we have
\begin{equation}\label{eq:lagraomega}
    \mathcal{L}^{\prime}(\varphi,\Omega,\chi(\Omega))=\mathcal{L}_{\text{kin}}(\varphi)+\frac{1}{2}\Omega G_r^{-1}\left(-\frac{\square}{M^2}\right)\Omega-\frac{1}{2}\Omega\varphi^2.
\end{equation}
The $\Omega$ propagator in momentum space is
\begin{equation}
iD_{\Omega}(p^2)=i G_{r}(p^2/M^2),
\end{equation}
while that of the field $\varphi$ is 
\begin{equation}
    iD_{\varphi}^{ab}(p)=\frac{i\delta^{ab}}{(p^2-m^2+i\epsilon)F_n(p^2/M^2)}\equiv i\delta^{ab}D_{\text{HD}}(p^2),
\end{equation}
where we assume that all the zeros of $F_n$ are real and each pole $M^2_i$ is shifted to $M_i^2-i\epsilon$. The function $D_{\text{HD}}$ can always be decomposed into a sum of simple fractions as
\begin{equation}\label{eq:hdpropdecomp}
    D_{\text{HD}}(p^2)=\frac{a_0}{p^2-m^2+i\epsilon}+\sum_{i=1}^{n}\frac{a_i}{p^2-M_i^2+i\epsilon}, \qquad \text{with} \qquad \sum_{i=0}^na_i=0,
\end{equation}
where $a_i$ are real constants. 

The tree-level amplitude for a generic process $\varphi^a\varphi^b\rightarrow\varphi^c\varphi^d$ is
\begin{equation}\label{eq:treeampG}
\mathcal{M}_{ab\rightarrow cd}=-G_r(s)
\delta ^{ab}\delta ^{cd}-G_r(t) \delta
^{ac}\delta ^{bd}-G_r(u) \delta ^{ad}\delta
^{bc},
\end{equation}%
where $s=(p_{a}+p_{b})^{2}$, $t=(p_{a}+p_{c})^{2}$, $u=(p_{a}+p_{d})^{2}$
are the Mandelstam variables. Depending on $F_n$ and $G_r$, the cross section of a given process can grow as a power of $s$.

In this section, we do not specify $F_n$ and $G_n$ and show that in general the presence of purely virtual particles changes the high-energy behavior of the self energies. Using~\eqref{eq:bubint}, the one-loop $\Omega$ self energy is given by $N \Sigma(p^2)$ where
\begin{equation}
    \Sigma(p^2)=\int\frac{\text{d}^Dq}{(2\pi)^D}D_{\text{HD}}(p+q)D_{\text{HD}}(q)=\sum_{i,j=0}^na_ia_jB_{ij}(p^2), \qquad M_0\equiv m.
\end{equation}
Using the modified bubble diagram~\eqref{eq:Btau} we define $\Sigma(p^2,\tau)$ as
\begin{equation}
    \Sigma(p^2,\tau)\equiv\sum_{i,j=0}^na_ia_jB_{ij}(p^2,\tau).
\end{equation}
We want to show how $\Sigma(s,\tau)$ is modified by the different prescriptions. More precisely, in the large-$s$ expansion the terms of order $\mathcal{O}(1/s^0)$ and $\mathcal{O}(1/s)$ are always absent in the imaginary part, while in the real part they cancel only in the ghost case. In order to see this we first write the large-$s$ expansion of $B_{ij}(s,\tau)$. From now on, we omit the divergent part of the bubbles, assuming that they are subtracted by suitable counterterms in the modified minimal subtraction scheme. The leading order in the large-$s$ expansion of $B_{ij}(s,\tau)$ is 
\begin{equation}
B_{ij}(s,\tau)=B_{ij}^{(1)}(s,\tau)+\mathcal{O}(1/s^2),
\end{equation}
where $B_{ij}^{(1)}(s,\tau)$ contains the terms $\mathcal{O}(1/s)$ and its real and imaginary parts read
\begin{equation}
\text{Re}B_{ij}^{(1)}(s,\tau)=-\frac{\mathcal{T}_{ij}(\tau)}{16 \pi}\left(1-\frac{M_i^2+M_j^2}{s}\right),
\end{equation}
\begin{equation}
    \text{Im}B_{ij}^{(1)}(s,\tau) = \text{Im}B_{ij}^{(1)}(s,1)=-\frac{1}{16\pi^2}\left[\ln\frac{s}{M_iM_j}-\frac{1}{s}\left(M_i^2+M_j^2-M_i^2\ln\frac{s}{M_i^2}-M_j^2\ln\frac{s}{M_j^2}\right)\right],
\end{equation}
respectively.
For simplicity, throughout the paper we include the terms $(\ln s)/s^n$ in $\mathcal{O}(1/s^n)$.

In the ghost case ($\tau=1$) the large-$s$ expansion of the $\Omega$ self-energy is
\begin{equation}
    -16 \pi \text{Re}\Sigma(s,1)=\sum_{i=0}^na_i\sum_{j=0}^na_j\left(1-\frac{2}{s}M_j^2\right)+\mathcal{O}(1/s^2),
\end{equation}
\begin{equation}
    -16 \pi^2 \text{Im}\Sigma(s,1)=\sum_{i=0}^na_i\sum_{j=0}^na_j\left[\ln\frac{s}{M_j^2}-\frac{2}{s}M_j^2\left(1-\ln\frac{s}{M_j^2}\right)\right]+\mathcal{O}(1/s^2).
\end{equation}
Hence, using that $\sum_{i=0}^na_i=0$,
\begin{equation}\label{eq:sigma1order}
    \Sigma(s,1)=\mathcal{O}(1/s^2).
\end{equation}
For the more general case we assume that the particles $M_i$ with $0\leq i\leq k$, $k\in\mathbb{N}^+$ are standard. Therefore, $\mathcal{T}_{ij}(\tau)$ is made by a block with a $k\times k$ matrix with all the entries equal to 1, while the entries in the other blocks are equal to $\tau$. Using the explicit form of $\mathcal{T}_{ij}$ and the symmetry in $i,j$, the real part of the self energy can be written as
\begin{equation}\label{eq:resigmatauexpanded}
\begin{split}
\text{Re}\Sigma(s,\tau)&=\text{Re}\sum_{i,j=0}^ka_ia_jB_{ij}(s,1)+\tau\text{Re}\left[2\sum_{i=0}^k\sum_{j=k+1}^na_ia_jB_{ij}(s,1)+\sum_{i,j=k+1}^na_ia_jB_{ij}(s,1)\right]\\
&=\text{Re}\sum_{i,j=0}^ka_ia_jB_{ij}(s,1)+\tau\text{Re}\left[\Sigma(s,1)-\sum_{i,j=0}^ka_ia_jB_{ij}(s,1)\right]\\
&=(1-\tau)\text{Re}\sum_{i,j=0}^ka_ia_jB_{ij}^{(1)}(s,1)+\mathcal{O}(1/s^2).
\end{split}
\end{equation} 
where in the last step we have expanded in powers of $1/s$ and used~\eqref{eq:sigma1order}.
The imaginary part is unchanged. From the general formula~\eqref{eq:resigmatauexpanded} we can read the ghost case ($\tau=1$) and the case where $n-k$ particles are purely virtual ($\tau=0$). Depending on $F_n$ and $G_r$, this difference can be crucial when the resummed $\Omega$ propagator is used to compute the cross sections. 

To prepare the ground for the next section, we make use of the large-$N$ expansion and resum all the diagrams of order $1/N$. In order to discuss the diagrammatics in the large-$N$ expansion it is sufficient to know that the vertex function $G_{r}$ is linear in the couplings, which we label $\lambda_i$. Then, we introduce the 't Hooft couplings $\tilde{\lambda}_i\equiv N\lambda_i$, so $G_r(s,\lambda_i)=G_r(s,\tilde{\lambda}_i)/N$, and consider the large-$N$ limit while keeping each $\tilde{\lambda}_{i}$ finite. In this parametrization it is easy to derive which diagrams contribute to the leading order in $1/N$ (see e.g.~\cite{Parisi:1975im, Rosenstein:1990nm}). In fact, each $\Omega$ propagator is proportional to $1/N$, while each closed $\varphi$ loop gives a factor $N$. For definiteness we choose to study the scattering $\varphi^a\varphi^a\rightarrow \varphi^b\varphi^b$ with $a\neq b$. In this way we have only $s$-channel diagrams, as well as a few less diagrams in the next-to-leading order. However, we still need to specify which degree of freedom is set on shell in the external states. In fact, having higher-derivative kinetic terms, each $\varphi^a$ describes $n+1$ degrees of freedom. From now on we assume that the on-shell external states always satisfy the equation $p^2=m^2$, where $p_{\mu}$ is the four-momentum associated to an external leg.   
\begin{figure}
    \centering
    \includegraphics[width=16cm]{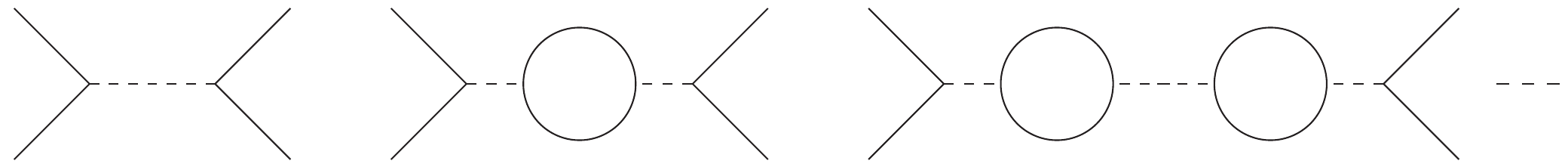}
    \caption{The leading diagrams in the $1/N$ expansion. The solid lines represent $\varphi$, while the dashed lines represent $\Omega$. The dashes on the right represent all other diagrams with an arbitrary number of self-energy insertions.}
    \label{fig:bubbleinsert}
\end{figure}
The tree-level diagram and the diagrams with an arbitrary insertion of self energies shown in figure~\autoref{fig:bubbleinsert} are all of order $1/N$. Such diagrams can be resummed and represented by a single tree-level diagram with the dressed propagator for $\Omega$, which reads
\begin{equation}\label{eq:resumomegaprop}
iD(s,\tau)=\frac{1}{N}\frac{iG_r(s,\tilde{\lambda}_i)}{1-i G_r(s,\tilde{\lambda}_i)\Sigma(s,\tau)}.
\end{equation}
Depending on the model, the presence of the first terms in~\eqref{eq:resigmatauexpanded} due to the fakeon prescription ($\tau=0$) leads to an improvement of the large-$s$ behavior of the cross sections. In the next section we show an explicit example. 

\sect{Explicit model}
\label{sect:super}
In this section we choose a particular theory of the class~\eqref{eq:generallagra} and show explicitly that the high-energy behavior of the cross section changes when we adopt the fakeon prescription. In particular, we choose
\begin{equation}
    n=r=1, \qquad F_1(z)=1-z, \qquad G_1(z)=\lambda_0-\lambda_1 z,
\end{equation}
where $\lambda_i$ are real parameters. The lagrangian reads
\begin{equation}\label{eq:superrenlagr}
\mathcal{L}(\varphi)=\frac{1}{2}(\partial _{\mu }\varphi^{a})\left( 1+\frac{\square 
}{M^{2}}\right) (\partial ^{\mu }\varphi^{a})-\frac{m^{2}}{2}\varphi
^{a}\left( 1+\frac{\square }{M^{2}}\right) \varphi^{a}-\frac{1}{8}\varphi^{2}\left( \lambda_0+\frac{\lambda_1 \square }{M^{2}}\right)\varphi^2,
\end{equation}%
where $M>m$. Such model has all the features necessary for our purposes. It contains $N$ standard particles of mass $m$ and $N$ ghosts of mass $M$. Moreover, it is (super)renormalizable (see~\autoref{app:reno}), while if we remove the higher-derivative kinetic term it becomes nonrenormalizable, pretty much like Stelle theory turns into Einstein gravity if the terms quadratic in the curvature are removed. Obviously, the case of gravity is more complex and requires a separate discussion. However, this simple model is enough to show some features which might be extended to the case of gravity.

The $\varphi$ propagator is
\begin{equation}
    iD_{\varphi}^{ab}(p^2)=-\frac{%
iM^{2}\delta^{ab}}{(p^{2}-m^{2}+i\epsilon)(p^{2}-M^{2}+i\epsilon)}
\end{equation}
and its decomposition~\eqref{eq:hdpropdecomp} is given by
\begin{equation}\label{eq:residues}
    M_0=m, \qquad M_1=M,\qquad a_0=-a_1=\frac{M^2}{M^2-m^2}>0.
\end{equation}
From~\eqref{eq:residues} we see that there are two degrees of freedom, one with positive and one with negative residues. The latter, if quantized using the Feynman prescription, is a ghost. Therefore, for the theory to satisfy the diagrammatic optical theorem we need $\tau=0$ for that degree of freedom, as explained in~\autoref{sect:unit}.

Another common feature of~\eqref{eq:superrenlagr} and Stelle gravity is that some tree-level cross sections grow with $s$. In the case of the process $\varphi^a\varphi^a\rightarrow\varphi^b\varphi^b$, with $a\neq b$, the cross section reads
\begin{equation}
\sigma(s)=\frac{a_0^2}{16\pi s}\left(\lambda_0-\lambda_1\frac{s}{M^2}\right)^2,
\end{equation}
where we have introduced the factor $a_0^2$ to account for the correct normalization of the phase space integral.
After introducing the auxiliary field $\Omega$ and following the steps of the previous section, the resummed cross section is given by
\begin{equation}\label{eq:crosssectD}
    \sigma(s,\tau)=\frac{a_0^2}{16 \pi s}|D(s,\tau)|^2.
\end{equation}
For $\tau\neq 1$ its expression at large $s$ and to the leading order in the large-$N$ expansion is
\begin{equation}\label{eq:sigmataularges}
    \sigma(s,\tau)=\frac{1}{(\tau-1)^2}\frac{16\pi}{a_0^2N^2}\left(\frac{1}{s}+\frac{4m^2}{s^2}\right)+\mathcal{O}(1/s^3),
\end{equation}
while for $\tau=1$ we have
\begin{equation}\label{eq:sigmaghostlarges}
    \sigma_{\text{gh}}(s)\equiv\sigma(s,1)=\frac{\tilde{\lambda}_1^2a_0^2}{16 \pi M^4 N^2}\left[s-\frac{\tilde{\lambda}_1a_0}{4 \pi^2}\left(m^2\ln\frac{s}{m^2}-M^2\ln\frac{s}{M^2}\right)\right]+\mathcal{O}(1/s^0).
\end{equation}
We see that the terms proportional to $1-\tau$ in the expansion~\eqref{eq:resigmatauexpanded} are crucial and give a different behavior in the cross section, which, after the resummation, decreases like $1/s$ for large $s$. On the other hand, the high-energy behavior in the case of ghosts is unchanged and the cross section still grows for large $s$, even after the resummation. 

Finally, we check the optical theorem for the resummed cross sections and show that it is violated in the ghost case, while it is satisfied in the PVP case. The optical theorem gives
\begin{equation}
    \text{Im}\left[-D(s,\tau)\right]=\sqrt{\kappa (s,m^2,m^2)}\sum_{X}\sigma_{X}(s,\tau),
\end{equation}
where $\sigma_{X}$ is the cross section of two $\varphi^a$ with mass $m$ going to the state $X$ and the sum runs over all possible final states. First, we note that every final state with more than two particles is subleading in the large-$N$ expansion, since each $\Omega$ propagator brings a factor $1/N$. Therefore, at the leading order, $X$ can be only a two-particle state, which include either states of mass $m$ or $M$. Then, 
since the external states have no spin and their masses are the same for any $\varphi^a$, the sum over $X$ gives $N$ copies of the same three possible cross sections. For $\tau=1$ we have
\begin{equation}
    \text{Im}\left[-D(s,1)\right]=N\sqrt{\kappa (s,m^2,m^2)}\left[\sigma_{\text{gh}}^{mm}(s)+2\sigma_{\text{gh}}^{mM}(s)+\sigma_{\text{gh}}^{MM}(s)\right],
\end{equation}
where
\begin{equation}\label{eq:optghost1}
       \sigma_{\text{gh}}^{m_1m_2}(s)=\frac{\sqrt{\kappa(s,m_1^2,m_2^2)}}{\sqrt{\kappa(s,m^2,m^2)}}\theta\left(s-(m_1+m_2)^2\right)\sigma_{\text{gh}}(s),
\end{equation}
while for $\tau=0$ we have
\begin{equation}\label{eq:optfake1}
    \text{Im}\left[-D(s,0)\right]=N\sqrt{\kappa (s,m^2,m^2)}\sigma(s,0).
\end{equation}
Note that the processes that involve particles of mass $M$ in the final states are removed by the fakeon prescription and, therefore, do not appear in~\eqref{eq:optfake1}.

It is easy to check whether equations~\eqref{eq:optghost1} and~\eqref{eq:optfake1} are satisfied by writing them in terms of the real and imaginary parts of $\Sigma$. In general, we have
\begin{equation}\label{eq:optSigma}
    \text{Im}\left[-D\right]=-\frac{1}{N}\frac{G_1^2\text{Re}\Sigma}{|1-iG_1\Sigma|^2},
\end{equation}
For $\tau=1$, \eqref{eq:optSigma} gives
\begin{equation}\label{eq:optghost2}
\begin{split}
    \text{Re}\Sigma(s,1)=-\frac{a_0^2}{16\pi}\Bigg[\sqrt{1-\frac{4m^2}{s}}\theta\big(s-&4m^2\big)+\sqrt{1-\frac{4M^2}{s}}\theta\left(s-4M^2\right)\\
    &+2\sqrt{\frac{\kappa(s,m^2,M^2)}{s^2}}\theta\left(s-(m+M)^2\right)\Bigg],
    \end{split}
\end{equation}
which is violated since the sign in front of $\sqrt{\kappa(s,m^2,M^2)}$  is opposite to that contained in $\text{Re}\Sigma(s,1)$. If we define the phase space such that $\sigma_{\text{gh}}^{mM}\rightarrow-\sigma_{\text{gh}}^{mM}$ then~\eqref{eq:optghost2} would be satisfied. However, we would have to deal with negative probabilities and negative cross sections, which have no physical meaning. 
\begin{figure}[t]
    \centering
    \includegraphics[width=11cm]{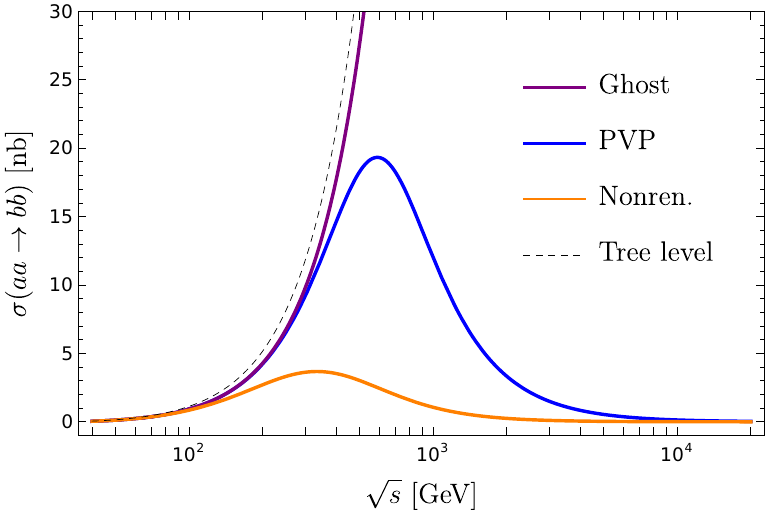}
    \caption{Resummed cross sections of the process $\varphi^a\varphi^a\rightarrow\varphi^b\varphi^b$ with $a\neq b$ as a function of the center-of-mass energy $\sqrt{s}$ for the ghost case (purple line), the PVP case (blue line) and the nonrenormalizable case (orange line), together with the tree-level amplitude (dashed line). The values of the parameters are $(m,M)=(15, 50)\text{GeV}$, $(\lambda_0,\lambda_1)=(0.1,0.3)$ and $N=1$.}
    \label{fig:PlotCrosSec}
\end{figure}

For $\tau=0$ we have
\begin{equation}\label{eq:optfake2}
    \text{Re}\Sigma(s,0)=-\frac{a_0^2}{16\pi}\sqrt{1-\frac{4m^2}{s}}\theta \left(s-4m^2\right),
\end{equation}
which is always satisfied.

For completeness, we also study the nonrenormalizable version of~\eqref{eq:superrenlagr}, i.e. without the higher-derivative kinetic term. In this case the cross section is given by
\begin{equation}
     \sigma_{\text{nr}}(s)=\frac{1}{16 \pi s}|D_{\text{nr}}(s)|^2, \qquad iD_{\text{nr}}(s)=\frac{1}{N}\frac{iG_1(s,\tilde{\lambda}_i)}{1-i G_1(s,\tilde{\lambda}_i)B_{00}(s,1)},
\end{equation}
and in the large-$s$ expansion we find  
\begin{equation}\label{eq:crosssectnr}
    \sigma_{\text{nr}}(s)=\frac{16 \pi}{s N^2\left(1+\ln^2\frac{s}{m^2}\right)}+\mathcal{O}(1/s^2),
\end{equation}
which also decreases. The quantitative difference between~\eqref{eq:crosssectnr} and~\eqref{eq:sigmataularges} is due to additional terms that appear in the imaginary part of $B_{00}(s,1)$, which is 
\begin{equation}\label{eq:nrimsigma}
    \text{Im}B_{00}(s,1)=-\frac{1}{16\pi^2}\ln\frac{s}{m^2}+\frac{m^2}{8\pi^2 s}\left(1+\ln\frac{s}{m^2}\right)+\mathcal{O}(1/s^2),
\end{equation}
in addition to the constant and $1/s$ terms in the real part, which are already enough to change the high-energy behavior. The fact that resummations of self energies can improve the behavior of cross sections in nonrenormalizable theories is known. For example, in~\cite{Parisi:1975im} the same procedure adopted here was used in the case of two-derivative $O(N)$ models, both in the bosonic and fermionic cases (in 5 and 3 dimensions, respectively). Moreover, also the case of general relativity has been studied in~\cite{Han:2004wt} and~\cite{Aydemir:2012nz}, where the scattering of massless scalars nonminimally coupled to gravity was considered and the graviton self energies (with scalar loops only) were resummed in the large-$N$ limit. Again, after the resummation the cross section satisfies unitarity. However, a treatment in the case of Stelle gravity is missing. We believe that the results of this paper suggest that resummations might cure the high-energy behavior of graviton scattering in Stelle gravity, provided that the ghost is turned into a PVP.

So far, we have used the large-$s$ expansion to facilitate the study of the high-energy behavior of the cross sections. However,~\eqref{eq:crosssectD} gives us the full formula for the cross sections at the leading order in $1/N$. Since the explicit formulas are rather cumbersome and not very instructive, we show only their plots, which give an idea of the modifications due to the resummation. They are depicted altogether in~\autoref{fig:PlotCrosSec}.
\begin{figure}[t]
    \centering
    \includegraphics[width=16cm]{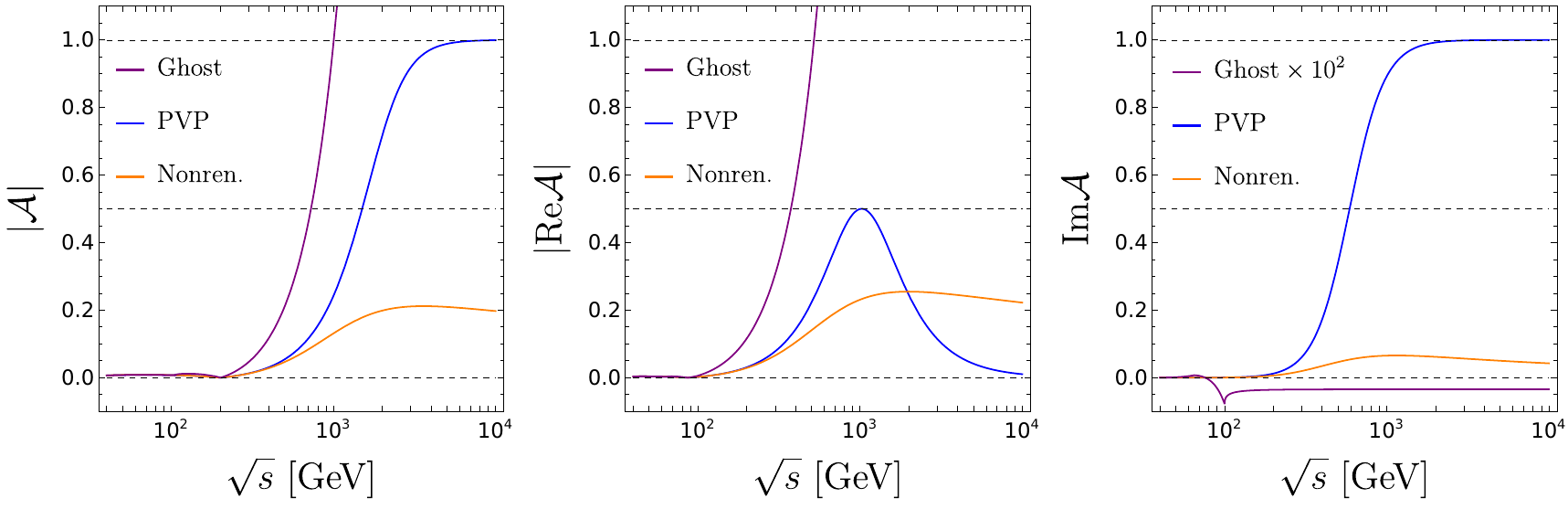}
    \caption{The modulus, modulus of real part and imaginary part of~\eqref{eq:Aghost} and~\eqref{eq:APVPandNR} as functions of the center-of-mass energy $\sqrt{s}$. The purple lines represent the ghost case, the blue lines represent the PVP case, while the orange lines represent the nonrenormalizable case. The horizontal dashed lines represent the bounds~\eqref{eq:uniteqA}. The values of the parameters are $(m,M)=(15, 50)\text{GeV}$, $(\lambda_0,\lambda_1)=(0.1,0.3)$ and $N=1$.}
    \label{fig:Plot3A}
\end{figure}
We can see that the PVP and nonrenormalizable cases (blue and orange curves, respectively) decrease at high energies and develop a peak at some finite value. These peaks correspond to poles of order one in the complex-$s$ plane and might be interpreted as new resonances. However, in the case considered in~\autoref{fig:PlotCrosSec} the orange curve has a rather large peak and a particle interpretation might be not applicable. On the other hand, the blue curve has a more pronounced peak and the narrow width approximation could be used there. However, this difference between the blue and the orange curves depends on the values of the masses and couplings and can even be switched by varying them. We think that the resummed cross sections should be studied in details as analytic functions of a complex variable in the context of more realistic models, such as quantum gravity, in order to give a physical interpretation.
This kind of study might give some bounds on the parameters of the theory, obtained by requiring that the resonance have positive mass squared and positive width (in those situations where the narrow width approximation can be applied). This goes beyond the purpose of this paper, but it is worth of investigation for the future.
 
To complete our analysis of unitarity, in~\autoref{fig:Plot3A} we show the quantities $|\mathcal{A}(s)|$, $|\text{Re}\left[\mathcal{A}(s)\right]|$ and $\text{Im}\left[\mathcal{A}(s)\right]$ where 
\begin{equation}\label{eq:Aghost}
    \mathcal{A}_{\text{gh}}(s)=-\frac{a_0^2}{16\pi}\left[\sqrt{1-\frac{4m^2}{s}}+2\sqrt{\frac{\kappa(s,m^2,M^2)}{s^2}}+\sqrt{1-\frac{4M^2}{s}}\right]D(s,1),
\end{equation}
\begin{equation}\label{eq:APVPandNR}
      \mathcal{A}_{\text{PVP}}(s)=-\frac{a_0^2}{16\pi}\sqrt{1-\frac{4m^2}{s}}D(s,0), \qquad  \mathcal{A}_{\text{nr}}(s)=-\frac{1}{16\pi}\sqrt{1-\frac{4m^2}{s}}D_{\text{nr}}(s)
\end{equation}
for the ghost, PVP and nonrenormalizable cases, respectively. Unitarity implies
\begin{equation}\label{eq:uniteqA}
    |\mathcal{A}(s)|\leq 1, \qquad \left|\text{Re}\left[\mathcal{A}(s)\right]\right|\leq \frac{1}{2}, \qquad 0\leq\text{Im}\left[\mathcal{A}(s)\right]\leq 1.
\end{equation}
From the purple lines in~\autoref{fig:Plot3A} we can see that~\eqref{eq:uniteqA} are violated in the case of ghost. In fact, both $|\mathcal{A}|$ and $|\text{Re}\mathcal{A}|$ diverge in the limit $s\rightarrow \infty$, while $\text{Im}A$ can be negative. The other two cases satisfy the bounds, as expected, since the diagrammatic optical theorem holds there. Note that in both~\autoref{fig:PlotCrosSec} and~\autoref{fig:Plot3A} we have set $N=1$ and the bounds are still satisfied. This might hint that the resummation can work also in pure gravity, where there is no analogous of $N$ to guide the resummation. In that case it could be enough to view the partial resummation as a prescription for asymptotic series, even without the help of a large-$N$ expansion.

To summarize, cross sections that grow as powers of $s$ in unitary theory can be modified by means of partial resummations. Therefore, such behavior does not signal the breakdown of predictivity or the need of new physics, but only the need of nonperturbative techniques. Nevertheless, nonrenormalizable theories remain predictive only at low energies due to the presence of infinitely many counterterms, each one with a new independent coupling. On the other hand, a renormalizable theory with PVP inherits only part of the behavior of its nonrenormalizable parent theory and, in the case of gravity, can be predictive well beyond the Planck scale.

\sect{Conclusions}
\label{sect:concl}
We have shown that resummations can cure apparent violations of unitarity bounds in a certain class of renormalizable theories with PVP, and in standard nonrenormalizable theories. We have pointed out how the behavior of the cross sections changes when PVP are introduced by studying the high-energy expansion of bubble diagrams. Moreover, using a higher-derivative $O(N)$ model in the large-$N$ limit, we resummed the leading order diagrams and obtained the full cross section, which has the correct decreasing behavior for high energies. A crucial point for such a mechanism to work is the validity of the diagrammatic optical theorem, that is to say the cutting equations can be interpreted as the diagrammatic expansion of the unitarity equation. Indeed, we have shown that renormalizable theories with ghosts cannot be cured in this way, and the presence of scattering cross sections that grow at high energies poses doubts on their predictive power, in contrast with the common believe that renormalizable theories can be predictive up to arbitrary high energies. 

Moreover, the unitarity bounds of the $O(N)$ model are satisfied even for $N=1$, which suggest that the same method might be used in pure gravity, where the situation in terms of scattering amplitudes is similar to that of the models studied in this paper, although without a large $N$. In that case, additional obstacles need to be overcome yet, such as the resummations of diagrams made by quartic vertices and handling their more complicated structure. We plan to deal with this in a future publication.

In the end, the fakeon prescription is able to retain not only the good properties of theories with ghosts that concern renormalizability, but also those of the parent nonrenormalizable theory that ensures good high-energy behavior in the finite parts.

Finally, the resummation introduces new complex poles in the cross sections that might be interpreted as resonances and deserve a deeper analysis.

\vskip.3 truecm \noindent {\large \textbf{Acknowledgments}}

\vskip .5 truecm

We are grateful to D. Anselmi, P. Baratella, L. Buoninfante and A. Melis for useful discussions.
\appendix
\renewcommand{\thesection}{\Alph{section}} \renewcommand{\theequation}{%
\thesection.\arabic{equation}} \setcounter{section}{0}

\sect{Renormalization}
\label{app:reno}
We derive the superficial degree of divergence for the theories studied in this paper, starting from~\eqref{eq:superrenlagr}. Switching to the variables $\hat{\varphi}^{a}=\varphi ^{a}/M$, $\hat{\lambda}_0%
=\lambda_0M^{4}$ and $\hat{\lambda}_1=\lambda_1 M^{2}$, we find%
\begin{equation}
\mathcal{L}=-\frac{1}{2}\hat{\varphi}^{a}\left( M^{2}+\square \right)
(m^{2}+\square )\hat{\varphi}^{a}-\frac{1}{8}\hat{\varphi}^{2}\left( 
\hat{\lambda}_0+\hat{\lambda}_1\square \right) \hat{\varphi}^{2},
\end{equation}%
up to a total derivative. The dimensions of fields and couplings are
\begin{equation}
\lbrack \hat{\varphi}^{a}]=0,\qquad \lbrack \hat{\lambda}_0]=4,\qquad \lbrack \hat{%
\lambda}_1]=2.
\end{equation}%
This shows that the counterterms are polynomials in couplings with positive dimensions. An $L$-loop diagram $\mathcal{D}$ with $I$ internal legs, $E$ external legs, $V_{0}$
vertices proportional to $\hat{\lambda}_0$ and $V_{1}$ vertices proportional to $%
\hat{\lambda}_1$ satisfies%
\begin{equation}
\omega (\mathcal{D})=4L-4I+2V_{1},
\end{equation}%
where $\omega (\mathcal{D})$ denotes the superficial degree of divergence. Using 
\begin{equation}\label{eq:diagrelations}
    L-I+V=1,\qquad 4V=E+2I,\qquad V=V_{0}+V_{1},
\end{equation}
we find%
\begin{equation}
\omega (\mathcal{D})=4-2V_{0}-2V,
\end{equation}
which is 2 for the one-loop tadpole with $V_0=0$ and 0 for $V_0=1$, while for all the other one-particle irreducible diagrams is smaller or equal than $0$, since the number of vertices $V$ cannot be smaller than 2. Thus, the only non-tadpole divergent
diagrams are logarithmically divergent and have $V=2$, $V_{0}=0$. There are
two such possibilities: $L=E=2$, $I=3$ and $
L=1 $, $E=4$, $I=2$. Therefore, the theory is two-loop superrenormalizable.

Adding the tadpole and switching back to the original variables, the one-loop renormalized lagrangian reads 
\begin{equation}\label{eq:renlagra}
\mathcal{L}_{\text{R}}=\frac{Z_{\varphi}}{2}\partial_{\mu}\varphi^a\partial^{\mu}\varphi^a-\frac{m^2Z_{m^2}}{2}\varphi^2-\frac{\lambda_0 Z_{\lambda_0}}{4!}(\varphi^2)^2,
\end{equation}
where
\begin{equation}
    Z_{\varphi}=1+\frac{\lambda_1}{8\pi^2 \varepsilon}+\mathcal{O}(\lambda_i\lambda_j), \qquad  Z_{\lambda_0}=1+\frac{\lambda_1^2}{4\pi^2\varepsilon}+\mathcal{O}(\lambda_i\lambda_j),
    \end{equation}
    \begin{equation}
   m^2Z_{m^2}=1+\frac{\lambda_1(m^2+M^2)}{8\pi^2\varepsilon}-\frac{\lambda_0M^2(N+2)}{16\pi^2\varepsilon}+\mathcal{O}(\lambda_i\lambda_j).
\end{equation}
After the resummation, the superficial degree of divergence changes, if we set $\tau=0$, while it stays the same for $\tau=1$. In fact, the new resummed vertex proportional to~\eqref{eq:treeampG} with the substitution
\begin{equation}
    G_r(z)\rightarrow \frac{1}{N}\frac{G_r(z,\tilde{\lambda}_i)}{1-i G_r(z,\tilde{\lambda}_i)\Sigma(z,\tau)},
\end{equation}
which for large $z$ and $\tau=0$ goes to a constant. Therefore, it does not contribute to the superficial degree of divergence, which now reads
\begin{equation}
    \omega (G)=4L-4I \qquad \Rightarrow\qquad \omega (G)=4-4V.
\end{equation}
Hence, every non-tadpole diagram ($V>1$) is finite. 

In the case of ghosts ($\tau=1$) the high-energy behavior of the vertex function is not modified and the superficial degree of divergence remains the same.

In general, the superficial degree of divergence for the lagrangian~\eqref{eq:generallagra} is
\begin{equation}
    \omega(\mathcal{D})=4L-2(n+1)I+2\sum_{k=0}^{r}kV_k,
\end{equation}
where $V_k$ is the vertex proportional to $\lambda_k$. Using again the relations~\eqref{eq:diagrelations} we can write
\begin{equation}\label{eq:generalomega}
\begin{split}
   \omega(\mathcal{D})&=4+(n-1)E-4nV+2\sum_{k=0}^{r}kV_k\\
   &\leq 4+(n-1)E-2(2n-r)V\\
   &=4-(n-r+1)E-2(2n-r)(L-1),
   \end{split}
\end{equation}
where in the last step we removed $V$ using~\eqref{eq:diagrelations} again. From~\eqref{eq:generalomega} we can see that some particular cases are superrenormalizable or finite. For example for $n=r>1$ we get
\begin{equation}
    \omega(\mathcal{D})\leq 4-E-2n(L-1),
\end{equation}
which shows that for $L=1$ only diagrams with $E=2,4$ are divergent, while for $L>1$ all the diagrams are finite and these theories are one-loop superrenormalizable. Note that in these cases the all the possible interactions of 4 fields and up to $2r$ derivatives cannot be all written in the form $\varphi^2G_r\varphi^2$. However, because of superrenormalizability, the renormalized lagrangian is always of the form~\eqref{eq:renlagra}. Therefore, there is no need to include all the other form of interaction, since they are not generated by renormalization.

Another possibility is to set $n>r\geq 1$. If $n-r=1$, then the only divergent diagram is the one-loop tadpole ($E=2$ and $L=1$), while for $n-r>1$ the theories are finite.

\bibliographystyle{JHEP} 
\bibliography{mybiblio}

\end{document}